\documentclass[12pt]{article}
\usepackage{times}
\usepackage{graphicx}
\usepackage{natbib}
\usepackage{color}

\newcounter{lastnote}

\title{Room temperature electrically tunable broadband terahertz Faraday effect}

\author
{Alexey Shuvaev$^{1}$, Andrei Pimenov$^{1\ast}$,
Georgy~V.~Astakhov$^{2}$, Mathias M\"{u}hlbauer$^{3}$\\
Christoph Br\"{u}ne$^{3}$, Hartmut Buhmann$^{3}$, Laurens W. Molenkamp$^{3}$ \\
\\
\normalsize{$^{1}$ Institute of Solid State Physics,
Vienna University of Technology, 1040 Vienna, Austria}\\
\normalsize{$^{2}$ Physikalisches Institut (EP6), Universit\"{a}t
W\"{u}rzburg, 97074 W\"{u}rzburg, Germany}\\
\normalsize{$^{3}$ Physikalisches Institut (EP3), Universit\"{a}t
W\"{u}rzburg, 97074 W\"{u}rzburg, Germany}\\
\\
\normalsize{$^\ast$To whom correspondence should be addressed;
E-mail:  pimenov@ifp.tuwien.ac.at} }

\date{}

\begin{document}

\baselineskip24pt

\maketitle

\textbf{The terahertz (THz) frequency range (0.1-10 THz) fills the
gap between the microwave and optical parts of the electromagnetic
spectrum. Recent progress in the generation
\cite{williams_nphot_2007} and detection \cite{wanke_nphot_2010} of
the THz radiation has made it a powerful tool for fundamental
research and resulted in a number of applications
\cite{tonouchi_nphot_2007, jepsen_lpr_2011}. However, some important
components necessary to effectively manipulate THz radiation are
still missing. In particular, active polarization and phase control
over a broad THz band would have major applications in science and
technology. It would, e.g., enable high-speed modulation for
wireless communications \cite{federici_jap_2010} and real-time
chiral structure spectroscopy of proteins and DNA
\cite{nafie_arpc_1997, xu_spie_2003}. In physics, this technology
can be also used to precisely measure very weak Faraday and Kerr
effects, as required, for instance, to probe the electrodynamics of
topological insulators \cite{maciejko_prl_2010, shuvaev_2012}. Phase
control of THz radiation has been demonstrated using various
approaches. They depend either on the physical dimensions of the
phase plate (and hence provide a fixed phase shift)
\cite{masson_ol_2006, strikwerda_oe_2009, saha_oe_2010} or on a
mechanically controlled time delay between optical pulses (and hence
prevent fast modulation) \cite{shimano_jjap_2005, amer_apl_2005,
dai_prl_2009}. Here, we present data that demonstrate the room
temperature giant Faraday effect in HgTe \cite{shuvaev_prl_2011} can
be electrically tuned over a wide frequency range (0.1-1 THz). The
principle of operation is based on the field effect in a thin HgTe
semimetal film. These findings together with {the low scattering
rate} in HgTe open a new approach for high-speed amplitude and phase
modulation in the THz frequency range. }

The idea behind our experiments is to exploit the strong interaction
between a solid-state plasma and THz radiation
\cite{wang_nphys_2010}. Because of this interaction, left- and
right-handed circularly polarized THz waves
have different complex refraction indices when a magnetic field is applied \cite{palik_rpp_1970}.
Recently, this characteristic property has been proposed to realize
broadband THz modulators and phase plates based on electron-doped
InSb crystals \cite{arikawa_oe_2012}. However, there are two
problems in the realization of such devices for practical applications. First, the
giant Faraday effect in InSb does not survive at room temperature.
Second, this approach requires fast modulation of the applied
magnetic field with an amplitude of several hundreds mT,
which is a rather challenging demand.

In semimetals with zero (or very small) band gap, such as HgTe, the
electron plasma exists in undoped samples, due to the thermal
activation of electrons. The ability to use such undoped HgTe layers
results in a very high electron mobility and, consequently, we have been
able to demonstrate the giant Faraday effect even at room temperature \cite{shuvaev_prl_2011}.

To achieve control over the Faraday effect in a constant magnetic field, we have now
fabricated devices fitted with transparent gate electrodes, which  allow to
change the carrier density and thus the properties of the electron plasma in the HgTe layer.  By applying a moderate
voltage $| U | < 1 \, \mathrm{V}$ to these gates we are able to control the Faraday
rotation and Faraday ellipticity, i.e., amplitude, phase shift, and
polarization of the THz radiation at room temperature.

\begin{figure}
\begin{center}
\includegraphics[angle=270, width=0.95\linewidth, clip]{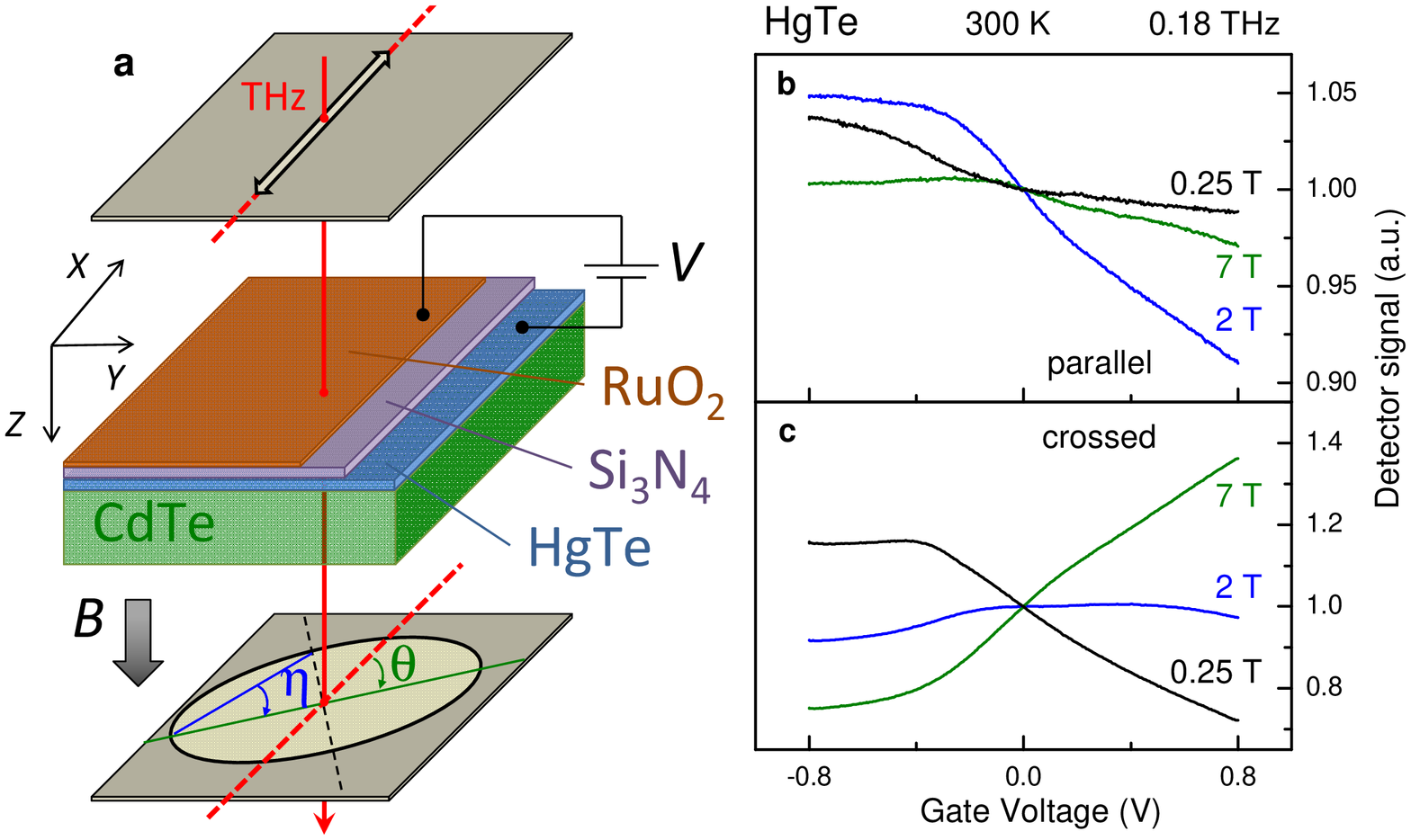}
\end{center}

\caption{\emph{Electric voltage control of THz radiation.}
(\textbf{a}) - Schematic representation of experimental arrangement
to control the Faraday rotation ($\theta$) and Faraday ellipticity ($\eta$) by a gate voltage. (\textbf{b,c}) -
Normalized detector signal as function of gate voltage in a
geometry with parallel (\textbf{b}) and crossed (\textbf{c})
polarizers, respectively.} \label{fig1}
\end{figure}

The THz experiments have been carried out in Faraday geometry as
shown schematically in Fig. 1\textbf{a}. In this geometry the
external magnetic field is parallel to the propagation direction of
the THz radiation. The incident polarization is linear and the
polarization of the transmitted radiation is detected by a wire grid
analyzer. Further experimental details are given in the
Supplementary Material. All experiments presented here have been
carried out at room temperature.  {Our estimations suggest that the
concentration of carriers confined to the topological surface state
in this material \cite{shuvaev_2012,hancock_prl_2011} is comparable
with the concentration of carriers occupied bulk states. }

As shown in Fig. 1 \textbf{bc}, we observe a substantial variation of
the amplitude of the transmitted signal as a function of gate
voltage, both in parallel ($t_p$) and in crossed ($t_c$) polarizer
geometry. Consequently, the Faraday rotation and ellipticity can be
controlled by using a moderate electric voltage. In a first
approximation, the observed changes may be attributed to the gate
dependence of the effective charge density in the HgTe film. As will
be shown below, not only the density, but all electrodynamic
parameters of the HgTe are affected by the gate voltage. We mention
already at this point that the response of the THz signal is asymmetric
with the sign of the gate voltage, which is probably due to carrier trapping
at the interface between HgTe and the gate insulator. We also observed
hysteresis effect when tuning the gate voltage beyond certain critical value.

\begin{figure}
\begin{center}
\includegraphics[ width=0.65\linewidth, clip]{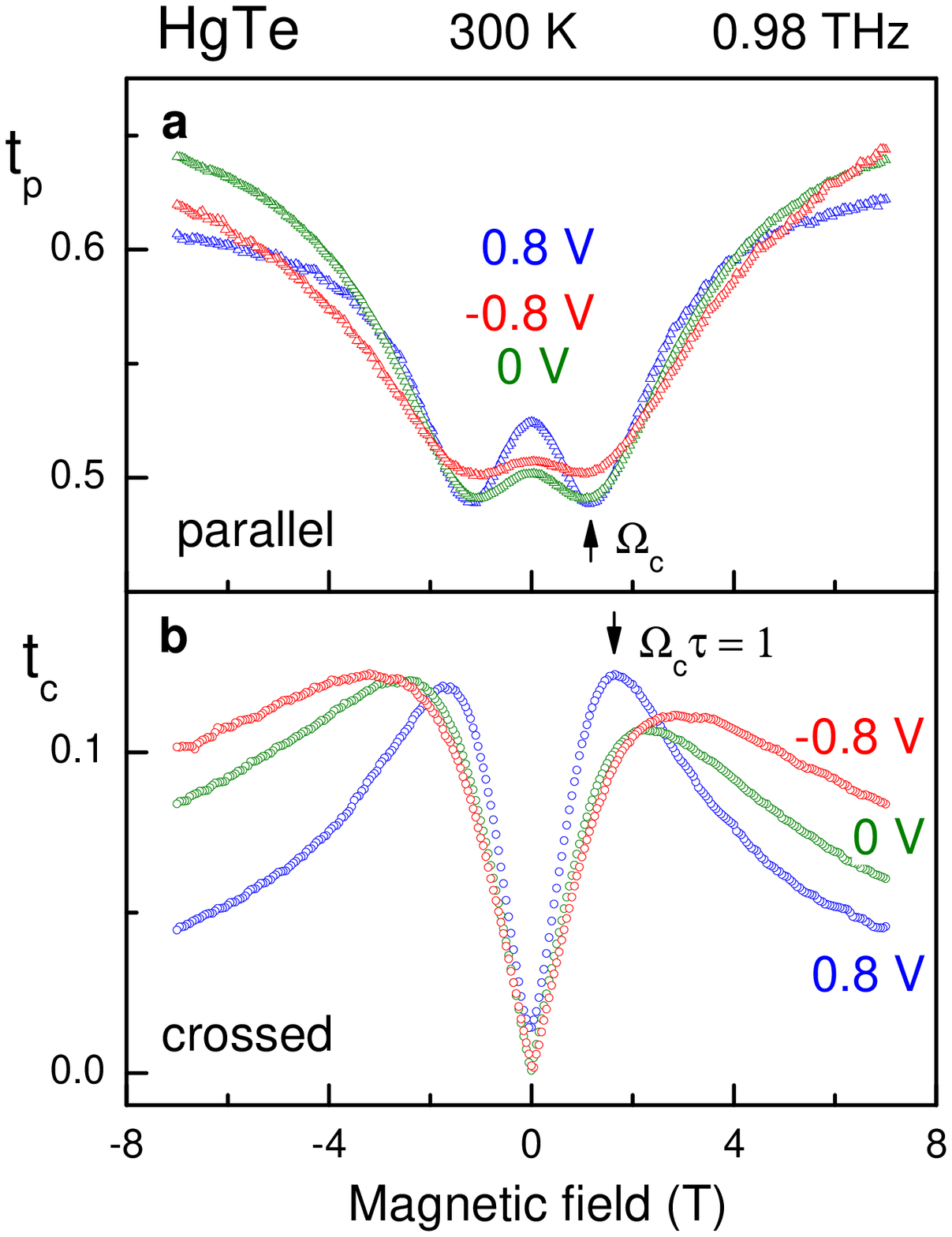}
\end{center}

\caption{\emph{Magnetooptic transmittance of the HgTe film.} Magnetic
field dependent amplitude of the transmitted radiation in Faraday
geometry and for different gate voltages as indicated. (\textbf{a}) -
Transmission in parallel polarizers geometry. The arrow marks the
position of the cyclotron resonance $\omega = \Omega_c$. (\textbf{b}) -
Transmission in crossed polarizers. The maximum indicated by the arrow
allows a determination of the quasiparticle scattering time $\tau$.}
\label{fig2}
\end{figure}

In order to characterize the electron dynamics in HgTe under applied
gate voltage, a set of experiments at different magnetic fields $B$ and
THz frequencies $\omega$ has been carried out. Figure 2 demonstrates the
magnetic field dependencies of the magnetooptical signal of the HgTe
film at three different gate voltages. The transmission amplitude
for parallel polarizers (Fig. 2\textbf{a}) shows a clear minimum at
a frequency-dependent field position which corresponds to the
cyclotron resonance in HgTe $\Omega_c = \omega$. The width of the minimum in
$t_p$ is directly connected to the scattering rate of electrons. As
the width in Fig. 2\textbf{a} changes strongly with gate voltage, a
variation of the characteristic scattering time $\tau$ is evident.

Figure 2\textbf{b} shows the amplitude of the transmission in
crossed polarizer geometry. All curves reveal a characteristic
maximum at a magnetic field around $\pm 2$ T. As shown in the
Supplementary material [Eq. (\ref{sxxs})], this maximum corresponds
to a field value of $\Omega_c\tau=1$ and it allows for an
independent determination of $\tau$. Again, a
strong variation of the maximum is observed in the field
dependence of $t_c$, which confirms the variation of the scattering time with gate voltage.

\begin{figure}
\begin{center}
\includegraphics[width=0.65\linewidth, clip]{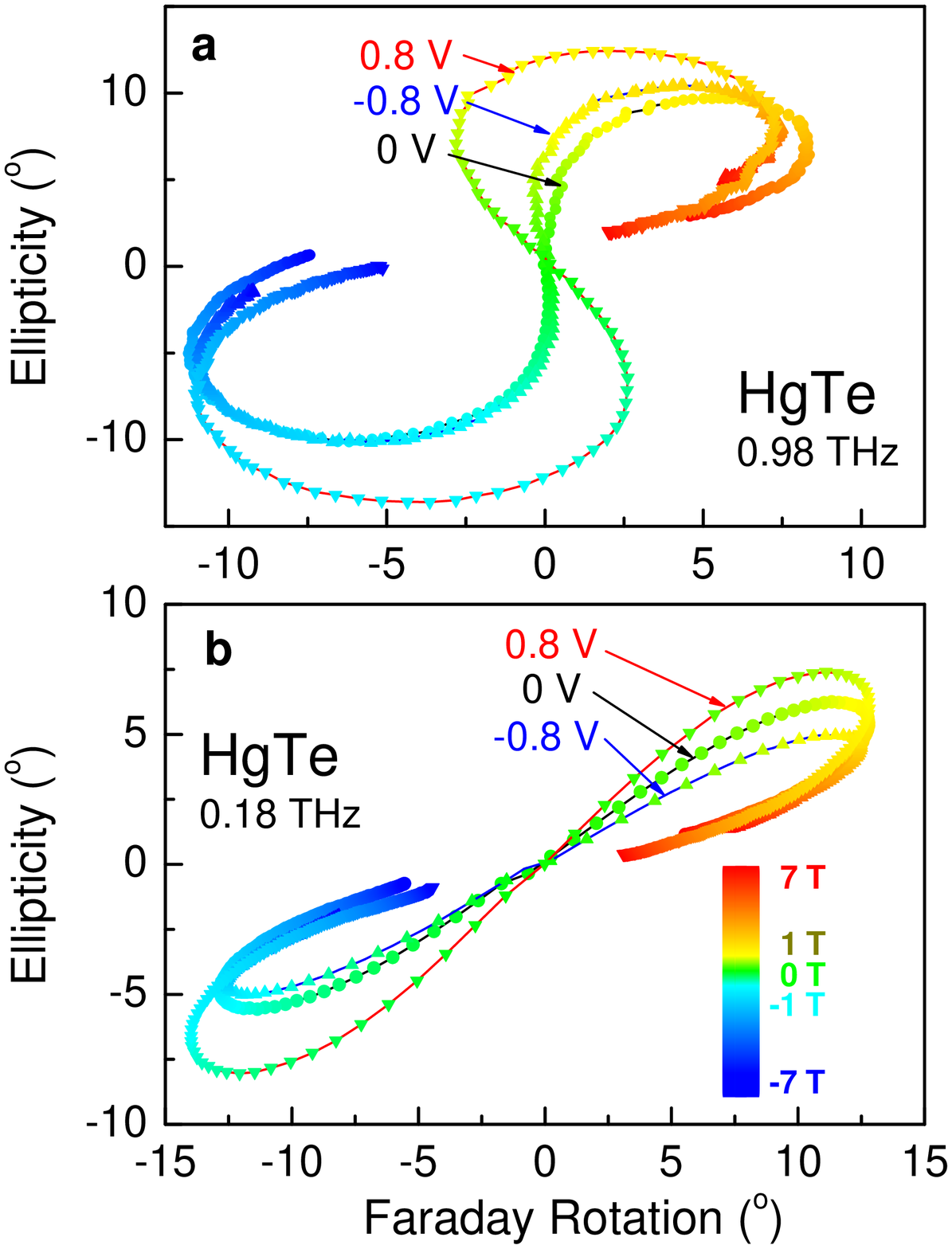}
\end{center}

\caption{\emph{Tunability of Faraday rotation and ellipticity.}
Parametric presentation of the attainable range of Faraday
rotation and ellipticity for different fields and voltages at 0.98
THz (\textbf{a}) and 0.18 THz (\textbf{b}), respectively. The data
are shown as magnetic field dependent polarization states at three
different gate voltages, as indicated. The values of the external
magnetic field are color-coded. Symbols - experiment, lines are a
guide to the eye.} \label{fig3}
\end{figure}

In addition to the transmission amplitude, phase
information has been obtained using a Mach-Zehnder interferometric
arrangement. From these experiments,  the polarization
state of the transmitted THz radiation can be fully polarized, including
Faraday rotation and ellipticity. These data are shown in Fig. 3
for the parameter range accessible in our setup ($| B | < \mathrm{7} \, T$ and
$ | U | < 0.8 \, \mathrm{V}$).
The data in Fig. 3 are given for two different frequencies, close to
the upper and the lower limits of our spectrometer. From the Faraday
ellipticity vs. Faraday rotation diagram of Fig. 3  one can see that
a desired polarization state can be set  by varying either magnetic
field or gate voltage.

\begin{figure}
\begin{center}
\includegraphics[width=0.75\linewidth, clip]{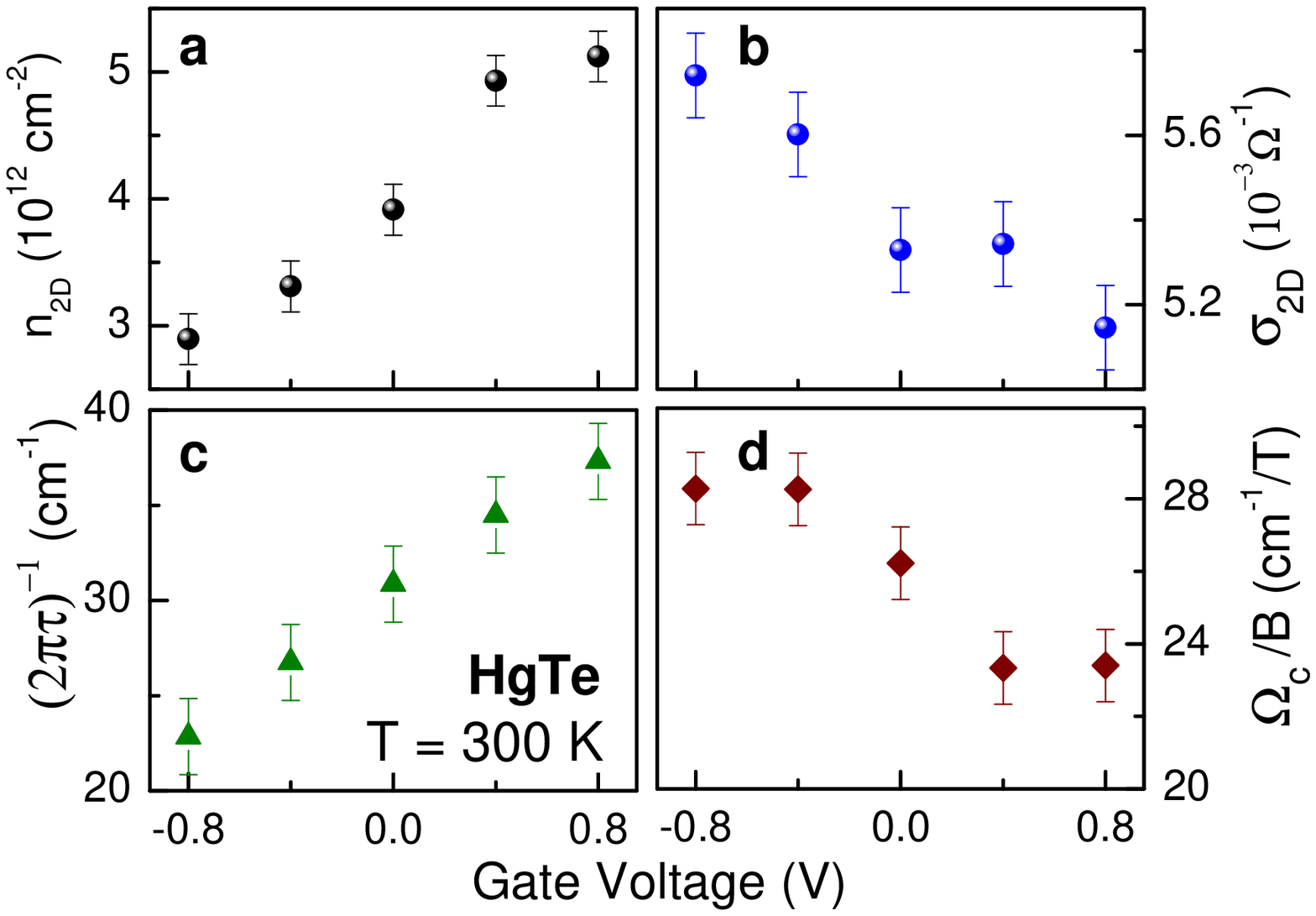}
\end{center}

\caption{\emph{Gate dependence of the electrodynamic parameters.}
Electrodynamic parameters of the conduction electrons in HgTe at
room temperature as obtained from Faraday rotation and
ellipticity. \textbf{a} - Effective two-dimensional electron
density. \textbf{b} - Two-dimensional $dc$ conductivity. \textbf{c}
- Electron scattering rate. \textbf{d} - Cyclotron resonance
frequency.} \label{fig4}
\end{figure}

The observed gate dependence of  the magnetooptical spectra can be
well interpreted within the Drude model. The details of the
characterization procedure and extraction of the electrodynamic
parameters are given in the Supplementary Material. In brief, we fit
simultaneously the magnetic field dependencies using charge density
($n_{2D}$), scattering rate ($1/\tau$) and {cyclotron frequency (in
terms of $\Omega_c / B$)} as the only fitting parameters. By
repeating this fit for different gate voltages, we determine how
these parameters depend on $U$, as shown in  Fig. 4.
Two-dimensional electron conductivity derived from
the other parameters as
\begin{equation}\label{eqsig}
{ \sigma_{2D}=   e n_{2D}  \frac{\Omega_c}{B}  \tau \, }
\end{equation}
is also shown in Fig. 4.

As expected, a strong variation of the electron density with gate
voltage (nearly a factor of two) is observed (Fig. 4\textbf{a}).
Assuming the simple capacitor formula $\Delta
n_{2D}=\varepsilon_0\varepsilon_{\rm{Si_3N_4}}\Delta
U/d_{\rm{Si_3N_4}}$ to estimate the density variation, we obtain
$\Delta n_{2D} \simeq 4.15 \cdot 10^{12}$ cm$^{-2}$ per Volt. Here
$\varepsilon_0$ is the permittivity of the free space and
$\varepsilon_{\rm{Si_3N_4}}=7.5$ is the dielectric constant of
$\rm{Si_3N_4}$ insulating layer. This is in a reasonable agreement
with our experimentally observed value of $1.12 \cdot 10^{12}$
cm$^{-2}$V$^{-1}$. Where again the difference with the capacitor
model likely results from carrier trapping at the
$\mathrm{HgTe/Si_3N_4}$ interface.

Contrary to the increase of the electron density with the gate
voltage, the conductivity ($\sigma_{2D}$, Fig. 4\textbf{b}) is a
decreasing function of $U$. This is rather unusual and can
conceivably be explained by increased carrier scattering
($\tau^{-1}$, Fig. 4\textbf{c}) at the $\mathrm{HgTe/Si_3N_4}$
interface for larger gate voltages. Finally, we observe that the
cyclotron frequency depends on the gate voltage as well, Fig.
4\textbf{d}. {This result is expected for the Dirac surface states
in HgTe with $\Omega_c / B \propto 1 / \sqrt{n_{2D}}$. It can be
also caused by the deviation from ideal parabolic dispersion of bulk
electrons in HgTe. }

All results presented here have been obtained at room temperature,
suggesting practical applications, such as fast phase and amplitude
modulation or direct control of the polarization state by gate voltage and/or
magnetic field. As the observed tuning of the THz radiation
is due to modulation of the electron conductivity in the HgTe layer and
is obtained by purely electrical means, it should be possible to achieve rather high
modulation speed, comparable to this obtained with high electron mobility transistors (HEMT).
The modulation amplitude is as large as several
degrees per Volt, and we believe that it can be further improved by
optimizing parameters of the gate material, insulating barrier and HgTe layer.


\subsection*{Acknowledgements}
This work was supported by the by the German Research Foundation DFG
(SPP 1285, FOR 1162) the joint DFG-JST Forschergruppe on
'Topological Electronics', the ERC-AG project '3-TOP', and the
Austrian Science Funds (I815-N16).

\hspace{2cm}
\newpage

\section*{Supporting Online Material}

The sample studied in this work is a coherently strained
100-nm-thick nominally undoped HgTe layer, grown by molecular beam
epitaxy on an insulating CdTe substrate \cite{becker_pss_2007}. A 3
nm thin $\mathrm{RuO_2}$ layer has been used as a gate electrode and 10 nm thick
Si$_3$N$_4$ layer as an insulating barrier. The gate electrode
transmits approximately 80 \% of the terahertz radiation and
revealed no own Faraday rotation signal. The amplitude of the gate
voltage has been limited to $\pm 0.8$ V to avoid an electrical
breakdown of the insulating layer.

Transmittance experiments at terahertz frequencies (0.1 THz $< \nu
<$ 1 THz) have been carried out in a Mach-Zehnder interferometer
arrangement~\cite{volkov_infrared_1985, pimenov_prb_2005} which
allows measurement of the amplitude and phase shift of the
electromagnetic radiation in a geometry with controlled
polarization. Using wire grid polarizers, the complex transmission
coefficient can be obtained both in parallel and crossed polarizers
geometry. Static magnetic fields, up to $\pm7$~Tesla, have been
applied to the sample using a split-coil superconducting magnet.

To interpret the experimental data we use the ac conductivity tensor
$\hat{\sigma} (\omega)$ obtained in the classical (Drude) limit from
the Kubo conductivity of topological surface states (see e.g. Ref.
\cite{tse_prb_2011}). The diagonal, $\sigma_{xx} (\omega)$, and
Hall, $\sigma_{xy} (\omega)$, components of the conductivity tensor
as functions of THz frequency $\omega$ can be written as:
\begin{eqnarray}
&& \sigma_{xx} (\omega)=\sigma_{yy} (\omega) = \frac{1-i \omega
\tau}{(1-i \omega \tau)^2 +(\Omega_c \tau)^2} \sigma_0
\,, \label{sxx}\\
&& \sigma_{xy} (\omega)=-\sigma_{yx} (\omega)= \frac{\Omega_c
\tau}{(1-i \omega \tau)^2 +(\Omega_c \tau)^2} \sigma_0 \,.
\label{sxy}
\end{eqnarray}
Here $\Omega_c = eBv_F/\hbar k_F$ is the cyclotron frequency,
$\sigma_0$ is the dc conductivity, $B$ is the magnetic field, $v_F$,
$k_F$, $e$, and $\tau$ are the Fermi velocity, Fermi wave-number,
charge, and scattering time of the carriers, respectively. For the
Dirac spin-helical surface states the Fermi wave-number depends on
the 2D carrier density, $n_{2D}$, through relation $k_F=\sqrt{4\pi
n_{2D}}$, with no spin degeneracy. For massive carriers the
cyclotron resonance frequency can be written as $\Omega_c = eB/m$,
where $m$ is the effective electron mass in the parabolic
approximation.

For a free standing film and neglecting any substrate effects, the
complex transmission coefficients in parallel ($t_p$) and crossed
($t_c$) polarizers geometry can be written as:
\begin{eqnarray}
&& t_p =\frac{4+2\Sigma_{xx}}
{4+4\Sigma_{xx}+\Sigma_{xx}^2+\Sigma_{xy}^2}  \,, \label{tp}\\
&& t_c =\frac{2\Sigma_{xy}}
{4+4\Sigma_{xx}+\Sigma_{xx}^2+\Sigma_{xy}^2}  \,. \label{tc}
\end{eqnarray}
Here $\Sigma_{xx}$ and $\Sigma_{xy}$ are effective dimensionless 2D
conductivities, defined as: $\Sigma_{xx}=\sigma_{xx}dZ_0$ and
$\Sigma_{xy}=\sigma_{xy}dZ_0$ with the HgTe film thickness
$d=100$\,nm and the vacuum impedance $Z_0 \approx 377\,\Omega$. In
order to take into account the influence of the substrate  a
transfer matrix formalism
\cite{berreman_josa_1972,shuvaev_epjb_2011,shuvaev_prl_2011} is
utilized. The electrodynamic properties of the CdTe substrate are
obtained in a separate experiment on a bare substrate.

According to Eq. (\ref{tp}), the transmittance in parallel
polarizers ($t_p$) depends mainly on $\Sigma_{xx}$. The minima in
$|t_p|$ roughly correspond to the cyclotron resonance energy and
they scale with magnetic field. This may be understood by taking into
account that for our film $\Sigma \ll 1$ and Eqs.
(\ref{tp},\ref{tc}) simplify to:
\begin{equation}\label{trsimple}
    t_p \simeq 1- \Sigma_{xx}/2; \quad t_c \simeq \Sigma_{xy}/2 \ .
\end{equation}
In the limit $\omega\tau \gg 1$, Eq. (\ref{sxx}) may be approximated
by
\begin{equation}\label{sxx1}
    \sigma_{xx} \simeq \frac{1-i \omega \tau}{(\Omega_c ^2-\omega ^2)\tau^2}
\sigma_0  \ ,
\end{equation}
which leads to a resonance like feature in $\sigma_{xx}$ and $t_p$
for $\Omega_c=\omega$. Thus, the positions and widths of the minima
in Fig. 2\textbf{a} are directly connected to Fermi velocity ($v_F$) for linear dispersion [or the effective mass
($m$) for parabolic dispersion] and the scattering rate ($\tau^{-1}$) of the electrons.

The polarization rotation $\theta$ and the ellipticity $\eta$ are
obtained from the transmission data using:
\begin{eqnarray}
&& \tan(2\theta)=2\Re(\chi)/(1-|\chi|^2)\ , \\
&& \sin(2\eta)=2\Im(\chi)/(1+|\chi|^2)\ .
\end{eqnarray}

Here $\chi=t_c/t_p$ and the definitions of $\theta + i \eta$ are
shown schematically in Fig. 1\textbf{a}. A direct interpretation of
the complex Faraday angle is in general not possible because of the
interplay of $\sigma_{xx}$ and $\sigma_{xy}$ in the data.

In the low frequency limit ($\omega\tau \ll 1$)  Eq. (\ref{sxy})
simplifies to the static result:
\begin{equation}\label{sxxs}
\sigma_{xy} = \frac{\Omega_c \tau}{1 +(\Omega_c \tau)^2}\sigma_0 \ .
\end{equation}
The last expression has a maximum at $\Omega_c (B)\tau = 1 $, which
leads to maxima in $t_c$ (Eqs. (\ref{tc}, \ref{trsimple}) and Fig.
2\textbf{b}) and in $\theta$ at about the same field value.
Therefore, the Faraday angle provides a direct and an independent
way of obtaining the scattering rate $1/\tau$.

\bibliography{literature_a}

\end{document}